\documentclass{ecai}

\usepackage{latexsym}
\usepackage{amssymb}
\usepackage{amsmath}
\usepackage{amsthm}
\usepackage{booktabs}
\usepackage{enumitem}
\usepackage{graphicx}
\usepackage{color}

\usepackage{xspace}
\usepackage[ruled,linesnumbered,vlined]{algorithm2e}
\usepackage{algpseudocode}
\usepackage{cleveref}
\usepackage{thm-restate}
\usepackage{multirow}
\usepackage[table]{xcolor}

\newtheorem{theorem}{Theorem}
\newtheorem{lemma}[theorem]{Lemma}

\newtheorem{definition}{Definition}

\newtheorem{example}{Example}

\makeatletter
\newcommand*\bigcdot{\mathpalette\bigcdot@{1.5}}
\newcommand*\bigcdot@[2]{\mathbin{\vcenter{\hbox{\scalebox{#2}{$\m@th#1\bullet$}}}}}
\makeatother
\usepackage{todonotes}

\newcommand{\A}{\mathcal{A}}
\newcommand{\B}{\mathcal{B}}

\newcommand{\D}{\mathcal{D}}

\newcommand{\F}{\mathcal{F}}

\newcommand{\M}{\mathcal{M}}
\newcommand{\N}{\mathcal{N}}

\newcommand{\T}{\mathcal{T}}

\newcommand{\bigO}{\mathcal{O}}

\newcommand{\MQ}{\mathsf{MQ}}
\newcommand{\EQ}{\mathsf{EQ}}

\newcommand{\wordletter}[2]{#1{[#2]}}
\newcommand{\subword}[3]{#1 {[#2..#3]}}

\newcommand{\alphabet}{\Sigma}
\newcommand{\emptyword}{\varepsilon}

\newcommand{\finwords}{\alphabet^*}
\newcommand{\infwords}{\alphabet^\omega}
\newcommand{\poswords}{\alphabet^+}
\newcommand{\langsymb}[0]{\mathcal{L}}
\newcommand{\lang}[1]{\langsymb(#1)}
\newcommand{\finlang}[1]{\langsymb_{*}(#1)}
\newcommand{\inflang}[1]{\langsymb(#1)}

\newcommand{\upword}[1]{\text{UP}(#1)}

\newcommand{\states}{Q}

\newcommand{\trans}{\delta}

\newcommand{\init}{q_0}

\newcommand{\run}{\rho}

\newcommand{\setnocond}[1]{\{#1\}}
\newcommand{\setcond}[2]{\{\, #1 \mid #2 \,\}}

\newcommand{\buchi}{B\"uchi\xspace}
\renewcommand{\@}{\xspace}

\newcommand{\size}[1]{|#1|}
\newcommand{\canoEq}{\backsim}

\newcommand{\class}[1]{[#1]_{\canoEq}}

\newcommand{\quotient}{\finwords/_{\canoEq}}

\newcommand{\BibTeX}{B\kern-.05em{\sc i\kern-.025em b}\kern-.08em\TeX}

\iffalse
	
	\newcommand{\sven}[1]{\todo[inline,color=teal!10,caption={Sven}]{\textbf{Sven:} #1}}
    \newcommand{\ly}[1]{\todo[inline,color=orange!10,caption={LY}]{\textbf{LY:} #1}}
	\newcommand{\mona}[1]{\todo[inline,color=green!10,caption={Mona}]{\textbf{Mona:} #1}}
 	\newcommand{\qiyi}[1]{\todo[inline,color=magenta!10,caption={Qiyi}]{\textbf{Qiyi:} #1}}
        \newcommand{\revise}[1]{\textcolor{blue}{#1}}
\else
        \newcommand{\revise}[1]{{\color{black}#1}}
	\newcommand{\sven}[1]{}
	\newcommand{\ly}[1]{}
	\newcommand{\mona}[1]{}
 	\newcommand{\qiyi}[1]{}

\fi

\begin{document}

\begin{frontmatter}

\paperid{123}

\title{Efficient Learning of Weak Deterministic B\"uchi Automata}

\author[A]{\fnms{Mona}~\snm{Alluwaym}%
}
\author[B]{\fnms{Yong}~\snm{Li}\orcid{0000-0002-7301-9234}\footnote{Corresponding author \{liyong@ios.ac.cn, qiyi.tang@liverpool.ac.uk\}}}
\author[A]{\fnms{Sven}~\snm{Schewe}\orcid{0000-0002-9093-9518}} 
\author[A]{\fnms{Qiyi}~\snm{Tang}\orcid{0000-0002-9265-3011}\footnotemark} 

\address[A]{University of Liverpool, Liverpool, UK}
\address[B]{Key Laboratory of System Software (Chinese Academy of Sciences) and State Key Laboratory of
Computer Science, Institute of Software, Chinese Academy of Sciences, Beijing, China}

\begin{abstract}
We present an efficient Angluin-style learning algorithm for weak deterministic B\"uchi automata (wDBAs). Different to ordinary deterministic B\"uchi and co-B\"uchi automata, wDBAs have a minimal normal form, and we show that we can learn this minimal normal form efficiently.
We provide an improved result on the number of queries required and show on benchmarks that this theoretical advantage translates into significantly fewer queries: while previous approaches require a quintic number of queries, we only require quadratically many queries in the size of the canonic wDBA that recognises the target language.
\end{abstract}

\end{frontmatter}

\section{Introduction}
\label{sec:intro}

This paper examines the $L^*$ learning framework introduced by Angluin~\cite{DBLP:journals/iandc/Angluin87}, commonly referred to as Angluin-style learning. This framework allows a learner to infer an automaton that represents an unknown system by interacting with a teacher through two types of queries: membership queries and equivalence queries. Membership queries ask whether a specific word $u$ belongs to the target language $L$, while equivalence queries ask whether a proposed automaton recognises $L$. After a sufficient number of membership queries, the learner constructs a conjectured automaton and submits it via an equivalence query.
The teacher either confirms that the conjecture accurately recognises $L$ or provides a counterexample, which the learner uses to locate the error and further refine the automaton. 
This process continues until the learner successfully converges to the correct automaton.

The Angluin-style learning framework has been applied in a variety of domains, including learning assumptions for compositional verification~\cite{DBLP:conf/tacas/CobleighGP03}, software testing~\cite{Czerny2014LearningbasedST}, detecting errors in network protocol implementations~\cite{DBLP:conf/uss/RuiterP15}, regular model checking~\cite{DBLP:conf/fmcad/ChenHLR17}, extracting automata models for recurrent neural networks~\cite{DBLP:conf/icml/WeissGY18}, verifying binarized neural networks~\cite{DBLP:conf/sat/ShihDC19}, and understanding machine learning models~\cite{DBLP:conf/ecai/Ozaki24}.

The foundation of the Angluin-style learning framework relies on the existence of a canonical form for the target automaton representation. For regular languages over \emph{finite} words, this is guaranteed by the Myhill-Nerode theorem~\cite{Myhill57,Nerode58}, which ensures that every regular language has a canonical minimal deterministic finite automaton (DFA). However, the situation is more complex for $\omega$-regular languages over \emph{infinite} words. It remains unknown whether deterministic B\"uchi and co-B\"uchi automata have minimal canonical forms, and the question is even more challenging for more powerful classes of $\omega$-automata, such as deterministic Rabin automata. As a result, learning automata over infinite words typically involves first learning a family of DFAs (FDFAs)~\cite{DBLP:journals/tcs/AngluinF16,DBLP:conf/atva/LiST23} and then converting the FDFA to the desired $\omega$-automaton~\cite{DBLP:journals/iandc/LiCZL21,LiST24}. This approach works because FDFAs have canonical forms for every $\omega$-regular language~\cite{DBLP:journals/tcs/AngluinF16,DBLP:conf/atva/LiST23}. However, such learning algorithms may suffer from exponential blow-up, as the canonical FDFA can be exponentially larger than the target $\omega$-automaton~\cite{LiST24}\footnote{A direct learning algorithm for $\omega$-automata exists~\cite{DBLP:journals/ai/MichaliszynO22}, which uses extra loop-index queries and thus does not fit into the Angluin-style learning framework.}. Nonetheless, a subclass of $\omega$-automata, known as \emph{weak} deterministic B\"uchi automata (DBAs), has canonical forms and can be directly learned~\cite{DBLP:journals/iandc/MalerP95}.

Weak DBAs (wDBAs) are an intriguing class of $\omega$-automata because they recognise weak languages—languages that can be recognised by both deterministic B\"uchi automata (DBAs) and deterministic co-B\"uchi automata (DCAs), or equivalently, languages that can express any Boolean combination of safety properties (ensuring that ``nothing bad happens") and reachability properties (ensuring that ``good things will happen"). 
For example, wDBAs have been used to express sets of assignments defined by formulas in additive or linear arithmetic over the reals and integers, i.e., $\langle \mathbf{R}, \mathbf{Z}, +, \leq \rangle $~\cite{DBLP:journals/tocl/BoigelotJW05}.
For example, the real $3.5$ can be encoded as an $\omega$-word $0^+11\cdot 10^{\omega}$ using binary representation. 
Vectors of reals can similarly be encoded over the alphabet $\{0, 1,\cdot\}^n$.
The set of satisfying assignments of a formula corresponds to a weak language~\cite{DBLP:journals/tocl/BoigelotJW05}.
This makes weak DBAs a highly expressive class of deterministic automata with a canonical normal form. Minimising wDBAs into this canonical form is tractable~\cite{DBLP:journals/ipl/Loding01}, whereas minimising even the next larger classes of automata, such as DBAs and DCAs, is already NP-hard~\cite{DBLP:conf/fsttcs/Schewe10}.

The current state-of-the-art learning algorithm for w%
DBAs remains the classic $L^\omega$ approach~\cite{DBLP:journals/iandc/MalerP95}. 
One might wonder whether it is possible to learn w%
DBAs using the efficient FDFA normal forms described in~\cite{DBLP:journals/tcs/AngluinF16} and~\cite{DBLP:conf/atva/LiST23}. 
However, these normal FDFAs are quadratic in size with respect to the minimal wDBA~\cite{DBLP:conf/atva/LiST23}; %
utilising them for learning is thus unlikely to improve over the state-of-the-art.

In this paper, we present a simple and novel wDBA learning algorithm that directly learns the canonical form, similar to $L^{\omega}$. 
\revise{Our main observation is that, by using an \emph{extra function} that stores a loop word for each state during the learning procedure, we can then easily determine whether a state is accepting and thus significantly reduce the queries needed.
This is in sharp contrast to the classic algorithm $L^{\omega}$~\cite{DBLP:journals/iandc/MalerP95}, which lacks this insight and has to compensate for it by looking for the loop words in the observation table each time. This causes asking redundant queries when building the conjectured wDBA.}
Notably, our algorithm offers several further improvements over the $L^\omega$ approach: it reduces the theoretical number of equivalence queries from quadratic to \emph{linear}, matching the theoretical bound in the finite-word setting. Additionally, we improve the theoretical number of membership queries from quintic to \emph{quadratic} in the size of the canonical wDBA. Unlike $L^\omega$, which heavily relies on the observation table to store query results, our algorithm is formulated independently of the data structure used for learning. This flexibility allows us to use observation tables from classic algorithms like $L^*$ \cite{DBLP:journals/iandc/Angluin87} as well as classification trees \cite{DBLP:books/daglib/0041035}, and it will allow us to benefit from future advances in data structures.

Our experimental results demonstrate that our approach requires significantly fewer queries compared to $L^\omega$.

\paragraph*{Related work.}

Learning algorithms for DBAs and DCAs are proposed in~\cite{LiST24}, where a DBA is learned via a limit family of DFAs (FDFA).
An FDFA consists of a \emph{leading} transition system to process the finite prefix $u$ and for each state of the transition system, there is a \emph{progress} DFA to accept the valid finite loops $v$.
This FDFA accepts $\omega$-words $uv^{\omega}$ by accepting the pair $(u, v)$.
The target limit FDFA can be quadratically larger than the minimal  wDBA~\cite{DBLP:conf/atva/LiST23}, making their learned DBA potentially \emph{quadratically larger} than ours.
Thus, our algorithm is at least quadratically more efficient in theory.  
In practice, \cite{LiST24} also requires significantly more queries than our algorithm on the selected benchmarks.

We use the binary analysis algorithm from~\cite{DBLP:journals/iandc/RivestS93} to analyse the counterexamples for the conjectured wDBA $\B$.
The classification tree data structure employed in this paper was originally proposed by~\cite{DBLP:books/daglib/0041035}.
A more advanced tree-based data structure called \emph{discrimination trees}~\cite{DBLP:conf/rv/IsbernerHS14} offers additional efficiency.
Integrating discrimination trees into our algorithm is a direction of future work.

Compared to $L^{\omega}$~\cite{DBLP:journals/iandc/MalerP95}, which relies heavily on observation tables to determine the acceptance of states, our approach introduces several key improvements.
First, we define an auxiliary function $g$ that maps each state to a loop word, enabling precise localisation of conflicts when a counterexample (CEX) is returned. Without this, $L^{\omega}$ must add all possible suffixes of the CEX to the observation table, resulting in a lot of membership queries to fill the entries of the table.
Our algorithm, by contrast, guarantees the addition of a new state after each equivalence query, whereas $L^{\omega}$ provides no such guarantee, with a worst-case bound of $n^2$ equivalence queries. This gap is not just theoretical---our experiments confirm the practical efficiency of our method.
Moreover, our algorithm is presented independently of specific data structures, allowing future adaptability.
We further employ global bounds on local parameters to reduce the number of iterations. Together, these design choices result in significantly fewer queries both theoretically and empirically.

\section{Preliminaries}
\label{sec:pre}
\label{sec:prelim}
In the whole paper, we fix a finite \emph{alphabet} $\alphabet$.
A \emph{word} is a finite or infinite sequence of letters in $\alphabet$;
$\emptyword$ denotes the empty word.
Let $\finwords$ and $\infwords$ denote the set of all finite and infinite words (or $\omega$-words), respectively.
In particular, we let $\poswords = \finwords\setminus\setnocond{\emptyword}$.
Let $\run$ be a sequence; 
we denote by $\wordletter{\run}{i}$ the $i$-th element of $\run$ and by $\subword{\run}{i}{k}$ the subsequence of $\run$ starting at the $i$-th element and ending at the $(k-1)$-th element when $0 \leq i < k$, and the empty sequence $\emptyword$ when $i \geq k$.
We denote by $\wordletter{\run}{i\ldots}$ the subsequence of $\run$ starting at the $i$-th element when $i < \size{\run}$, and the empty sequence $\emptyword$ when $i \geq \size{\run}$.
Given a finite word $u$ and a word $w$, we denote by $u \cdot w$ (or $uw$, for short) the concatenation of $u$ and $w$. 

\paragraph{\bf Transition system.}
A transition system (TS) is a tuple $\T = (\states, \init, \trans)$, where $\states$ is a finite set of states, $\init \in \states$ is the initial state, and $\trans: \states \times \alphabet \rightarrow \states$ is a transition function.
We also extend $\trans$ to words in a usual way, by letting $\trans(q, \emptyword) = q$ and $\trans(q, u \cdot a) = \trans(\trans(q, u), a)$, where $u \in \finwords$ and $a \in \alphabet$.
For simplicity, we denote by $\T(u)$ the state $\trans(q_0, u)$ without mentioning its transition function $\trans$ and the initial state $q_0$.

\paragraph{\bf Automata.}
An automaton on finite words is called a \emph{deterministic finite automaton} (DFA).
A DFA $\A$ is formally defined as a tuple $(\T, F)$, where $\T$ is a TS and $F\subseteq \states$ is a set of \emph{accepting} states.
An automaton on $\omega$-words is called a \emph{deterministic \buchi automaton} (DBA).
A DBA $\B$ is also represented as a tuple $(\T, F)$ like DFA.

The \emph{run} of a DFA $\A$ on a finite word $u$ of length $n \geq 0$ is a sequence of states $\run = q_{0} q_{1} \cdots q_{n} \in \states^{+}$ such that, for every $0 \leq i < n$, $q_{i+1} = \trans(q_{i}, \wordletter{u}{i})$.
A finite word $u \in \finwords$ is \emph{accepted} by a DFA $\A$ if its run $q_{0} \cdots q_{n}$ over $u$ ends in a final state $q_{n} \in F$. 
Similarly, the \emph{$\omega$-run} of $\A$ on an $\omega$-word $w$ is an infinite sequence of states $\run = q_{0} q_{1}\cdots$ such that, for every $i \geq 0$, $q_{i+1} = \trans(q_{i}, w[i])$. 
Let $\inf(\run)$ be the set of states that occur infinitely often in $\run$.
An $\omega$-word $w \in \infwords$ is \emph{accepted} by a DBA $\A$ if its $\omega$-run $\run$ of $\A$ over $w$ satisfies that $\inf({\run}) \cap F \neq \emptyset$. 
The language recognised by a DFA $\A$, denoted $\finlang{\A}$, is the set of finite words accepted by $\A$.
Similarly, we \linebreak denote by $\inflang{\A}$ the language recognised by a DBA $\A$.

A set $C\subseteq Q$ is said to be a \emph{strongly connected component} (SCC) if, for every two different states $p, q\in C$, we have $q \in \trans(p, u) $ and $p \in \trans(q, v)$ for some $u, v \in \poswords$.
A DBA is called \emph{weak} iff every %
SCC contains either only accepting
states or only rejecting states.
Therefore, every $\omega$-run $\rho$ of a weak DBA (wDBA) will eventually get trapped within either an accepting SCC or a rejecting SCC, i.e., $\inf({\rho}) \subseteq F$ or $\inf({\rho})\cap F = \emptyset$.
We say $\inf({\rho})$ is an accepting loop of $\rho$ if $\inf(\rho) \subseteq F$ and a rejecting loop if $\inf({\rho})\cap F = \emptyset$.

\section{Learning wDBA using RCs}
\label{sec:wDBA-learning}
\emph{Right congruences} (RCs) are at the core of the Angluin-style learning framework to distinguish finite word representatives of the states in the target automaton.
An RC relation is an equivalence relation $\canoEq$ over $\finwords$ such that $x \canoEq y$ implies $xv \canoEq yv$ for all $v \in \finwords$.
For a word $x \in \finwords$, we denote by $\class{x}$ the equivalence class of $\canoEq$ that $x$ resides in.
We denote by $\quotient$ the set of equivalence classes of $\finwords$ under $\canoEq$.

An RC $\canoEq$ defines a TS $\T$ where each state of $\T$ corresponds to an equivalence class in $\quotient$.
The TS $\T[\canoEq]$ is defined as follows.
\begin{definition}[\cite{Myhill57,Nerode58}]
\label{def:induced-dfw}
    Let $\canoEq$ be an RC with finite number of equivalence classes.
    The TS $\T[\canoEq]$ induced by $\canoEq$ is a tuple $(S, s_{0}, \trans)$ where $S = \quotient$, $s_{0} = \class{\emptyword}$, and for each $u \in \finwords$ and $a \in \alphabet$, $\trans(\class{u}, a) = \class{ua}$.
\end{definition}

The RC relation $\canoEq_R$ of a regular language $R$ is defined as: $u_1 \canoEq_R u_2$ iff for all $v\in \finwords$, we have $u_1v \in R \Leftrightarrow u_2v\in R$, where $u_1, u_2 \in \finwords$.
In fact, Angluin~\cite{DBLP:journals/iandc/Angluin87} uses $\canoEq_R$ to discover the word representative for each state in the TS $\T[\canoEq_R]$ and put all equivalence classes $[u]_{\canoEq_R}$ with $u \in R$ to the final state set $F$. 
In this way, every equivalence class of $\canoEq_R$ corresponds to a state in the minimal DFA of $R$.

For $\omega$-regular languages, however, the situation is more involved.
The first challenge is how to represent the infinite words as some word, like $ab^1 a b^2 \cdots ab^k \cdots $ with $k$ increasing to $\infty$, cannot be stored with finite memory.
The observation is that when dealing with an $\omega$-regular language $L$, we do not need to consider all types of $\omega$-words but a type of $\omega$-words called \emph{ultimately periodic} (UP) words $w$ that are in the form $uv^{\omega}$, where $u \in \finwords$ and $v \in \poswords$~\cite{Buc62}.

Naturally, a UP-word $w = uv^{\omega}$ can be represented as a pair of finite words $(u, v)$, called a \emph{decomposition} of $w$.
For an $\omega$-language $L$, let $\upword{L} = \setcond{uv^{\omega} \in L}{u \in \finwords \land v \in \poswords}$ denote the set of all UP-words in $L$.

\revise{
\begin{theorem}[\cite{Buc62}]
    Let $L$ and $L'$ be two $\omega$-regular languages.
    Then $L = L'$ holds if, and only if, $\upword{L} = \upword{L'}$.
\end{theorem}
Therefore, every $\omega$-regular language is uniquely characterised by its UP-words. Consequently, to learn an automaton of an $\omega$-regular language $L$, we only need to learn an automaton whose UP-words is $\upword{L}$.}

Moreover, unlike other $\omega$-automata, we have a canonical form of wDBAs for each wDBA language $L$~\cite{DBLP:journals/jcss/Staiger83}.
There is also an RC $\canoEq_L$ that induces the TS of the minimal wDBA of $L$, where the RC $\canoEq_L$ is defined as: for $u_1, u_2 \in \finwords$, 
\[ u_1 \canoEq_L u_2 \text{ iff }\forall (x\in\finwords, y \in \poswords). u_1 xy^{\omega} \in L \Leftrightarrow u_2 xy^{\omega} \in L.\]

\begin{figure}
    \centering
    \includegraphics[width=1\linewidth]{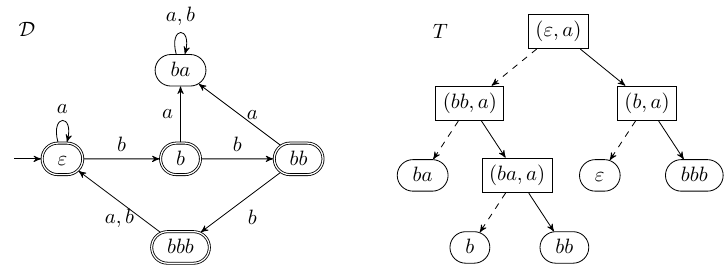}
        \caption{An example wDBA $\D$ and a classification tree $T$.}
    \label{fig:wdba-example}
\end{figure}

\begin{example}\label{ex:tree}
    Figure~\ref{fig:wdba-example} gives an example wDBA $\D$, which 
    accepts the language $ (a^* \cdot bbb\cdot (a | b))^{\omega} + (a^* \cdot bbb\cdot (a | b))^{*}a^{\omega}$.
    The states marked with double circles indicate that they are accepting states.
    On the right hand side, we depict a classification tree $T$ where each leaf node corresponds to an equivalence class of $\canoEq_L$ with respect to $\lang{\D}$.
    \revise{All internal nodes in the tree represent distinguishing
words such that a pair $(x, y)$ represents an $\omega$-word $xy^{\omega}$; they are used to distinguish the word representatives in leaf nodes.}
    Dashed arrows indicate a membership query result of $\bot$, while solid arrows represent a result of $\top$.
    For example, $\epsilon$ and $bbb$ can be distinguished by the node $(b, a)$. The transition from $(b, a)$ to $\epsilon$ is represented by a dashed arrow, indicating that $\epsilon \cdot ba^{\omega} \notin \lang{\D}$. 
    Conversely, the transition from $(b, a)$ to $bbb$ is shown with a solid arrow, indicating that $bbb \cdot ba^{\omega} \in \lang{\D}$.
\end{example}

\begin{figure}

    \centering
    \label{fig:wdba-table}
    \includegraphics[width=0.8\linewidth]{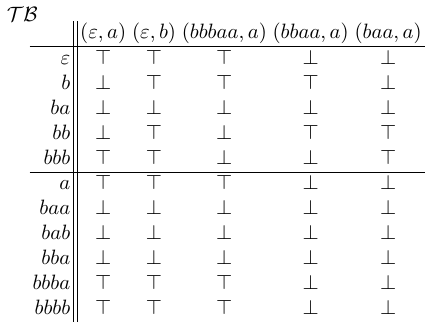}
    \caption{The observation table $\mathcal{TB}$ for the wDBA $\D$. We omit the columns for the suffixes $(aa, a)$ and $(a, a)$ because the entry values are the same as column $(\emptyword, a)$.}
\end{figure}

According to~\cite{DBLP:journals/iandc/MalerP95}, the canonical equivalence relation $\canoEq_L$ can be used to learn the TS of the minimal wDBA for a language $L$. Furthermore, the set of accepting states is determined by leveraging the samples stored in a data structure known as the observation table~\cite{DBLP:journals/iandc/Angluin87}. As an initial step, the experiment word $c^{\omega}$ is added to the columns for each letter $c \in \alphabet$.
Moreover, upon receiving a counterexample $w$ from the oracle, their algorithm adds all suffixes of $w$ to the columns of the observation table. 
A suffix of a UP-word $uv^{\omega}$ includes all words of the form $xv^{\omega}$ and $v'^{\omega}$, where $x$ is a prefix of $u$ and $v'$ is a rotation of $v$. 
For instance, the set of suffixes of $ad \cdot (bc)^{\omega}$ includes $ad \cdot (bc)^{\omega}$, $d \cdot (bc)^{\omega}$, $(bc)^{\omega}$, and $(cb)^{\omega}$.
Figure~\ref{fig:wdba-table} illustrates an observation table used by~\cite{DBLP:journals/iandc/MalerP95} to learn the wDBA $\D$ shown in Figure~\ref{fig:wdba-example}.
The entry for a row $u$ and column $(x, y)$ corresponds to the membership query result for the decomposition $(ux, y)$. 
In the observation table $\mathcal{TB}$, the columns include the words $a^{\omega}$ and $b^{\omega}$, as $\alphabet = \{a, b\}$.
For the counterexample $w = (bbbaa, a)$ \revise{returned by the teacher}, all suffixes of $w$ are added, including $(bbbaa, a)$, $(bbaa, a)$, $(baa, a)$, and $(aa, a) = (\epsilon, a)$.
The classification tree $T$ offers a more compact representation for storing membership query results compared to $\mathcal{TB}$. 
However, due to the suffix-adding operation, their algorithm heavily relies on observation tables, as it remains unclear how to apply this operation directly to classification trees~\cite{DBLP:books/daglib/0041035}.
In contrast, one advantage of our algorithm is that our algorithm can be implemented using either observation tables or classification trees.

Let $\T$ be the TS of a conjecture wDBA $\B$.
We say a decomposition $(u, v)$ is \emph{normalised} if $\T(u) = \T(uv)$.
That is, over the word $v$, $\T$ goes back to the state $\T(u)$.
Moreover, we denote by $\T_u(v)$ the state $\T(uv)$.

\section{Our wDBA learning algorithm}
\label{sec:wdba-learning}
We first give an overview of our wDBA learning algorithm and then present more details afterwards.
Assume that we have a wDBA oracle who knows $L$ and can answer membership queries about $L$ and equivalence queries about whether a given wDBA recognises $L$.

Our main idea is to use a TS learner to first learn the TS $\T$ of the target wDBA of $L$ and then use membership queries to decide whether a state in the TS belongs to the accepting state set $F$ or not.
Then, we obtain our wDBA $\B = (\T, F)$.
To this end, our wDBA learner is comprised of three components: the TS learner that learns a TS $\T$ (cf.\ \cref{sec:ts-learner}), the acceptance marking component that decides whether a state belongs to $F$ (cf.\ \cref{sec:acceptance-marking}), and a counterexample (CEX) analysis component (cf.\ \cref{sec:cex-analysis}).

Our learning algorithm starts by using a TS learner to learn a TS $\T$ via membership queries.
However, a challenge arises because the learned TS $\T$ lacks an acceptance mechanism $F$, which will then be addressed by our acceptance marking component.
To decide whether a state $u$ in $\T$ belongs to $F$ or not, we can just find a loop word $v \in \poswords$ such that $(u, v)$ is a \emph{normalised} decomposition in $\T$.
We mark the state $u$ as accepting or rejecting, i.e., put $u$ in $F$ or not, depending on the result of the membership query $\MQ(u, v)$.
The challenge here is that it is possible to find two different states within the same SCC marked with different acceptance labels.
This violates the definition of wDBAs, which requires all states within the same SCC to have a single acceptance label.
In this case, we say that we find a \emph{conflict} in $\T$.
To build a conjecture wDBA $\B$, we need to first make sure that $\T$ is \emph{conflict-free}.
We use the idea presented in~\cite{DBLP:journals/iandc/MalerP95} to resolve the conflicts.
Once no further conflicts have been found, we are able to construct a conjecture wDBA $\B = (\T, F)$ and ask an equivalence query about $\B$.

If the oracle returns ``yes" to our equivalence query, it means that our learning task has completed and we can output $\B$ as the correct wDBA of $L$.
In case that we receive the answer ``no" together with a counterexample $w \in \lang{\B} \ominus L$, the symmetric difference of $\lang{\B}$ and $L$, we will call the counterexample analysis component to obtain a valid counterexample for refining current conjecture $\B$.
In contrast to that the algorithm by Maler and Pnueli~\cite{DBLP:journals/iandc/MalerP95} does \emph{not} guarantee to increment the number of states in the conjecture by one after each counterexample guided refinement, our algorithm makes sure of that.
This then allows our algorithm to use at most $n$ equivalence queries instead of $\bigO(n^2)$ in~~\cite{DBLP:journals/iandc/MalerP95}, where $n$ is the number of states in the target wDBA $\D$. 

We will describe each component of the wDBA learner separately with more details in subsequent sections.

\subsection{The TS Learner}
\label{sec:ts-learner}
We will describe in this section the learner of the TS $\T$.

In~\cite{DBLP:journals/iandc/MalerP95}, Maler and Pnueli utilise an \emph{observation table}~\cite{DBLP:journals/iandc/Angluin87} to store the membership queries during the learning procedure.
To make our algorithm more general, we do not use a concrete data structure such as an observation table, a classification tree~\cite{DBLP:books/daglib/0041035} or other possible data structures to store membership queries.
Instead, we only assume that our learner has a function $f: S \times E  \cup S\cdot \alphabet\times E \rightarrow \{\top, \bot\} $, where $S \subseteq \finwords$ is a set of word representatives of the equivalence classes (i.e., the state names of the constructed TS), $E\subseteq \finwords\times \poswords$ is a set of experiments (UP-words) used to distinguish the words in $S$ and $\{\top, \bot\}$ is the function codomain.
Intuitively, for two different states/representatives $s_1, s_2 \in S$, we must have an experiment word $e \in E$ to distinguish $s_1$ and $s_2$, i.e., $f(s_1, e) \neq f(s_2, e)$ according to the RC $\canoEq$ we are using.

We let $\MQ(x, y)$ be the result of the membership query of $x\cdot y^{\omega}$ returned by the oracle.
The learning procedure begins by asking membership queries to define the function $f$ over its domain and then constructing a conjecture automaton for asking an equivalence query.
Note that $\emptyword \in S$ initially.

By the definition of $\canoEq_L$, for every $u \in S, (x, y) \in E$, we define $f(u,(x, y))  = \MQ(u \cdot x, y)$, i.e., the membership result of $u\cdot xy^{\omega}$.
In particular, for two different word representatives $u_1, u_2 \in S$, there must exist $(x, y) \in E$ such that $f(u_1, (x,y)) \neq f(u_2, (x, y))$, which means that $x\cdot y^{\omega}$ distinguishes the finite words $u_1$ and $u_2$ according to $\canoEq_L$.
When constructing $\T$, computing the $a$-successor of a representative $u$ reduces to finding a word representative $u' \in S$ such that $f(u', (x, y)) = f(ua, (x, y))$ for all $(x, y) \in E$.
Such a representative $u'$ will be guaranteed to exist during construction.
Note that $f(ua, (x, y))$ is also defined since the set $S\cdot \alphabet \times E$ is also part of the domain of $f$. 
The initial state will always be the state $\emptyword$.
This way, we obtain the TS $\T$.

Note that, for each $u \in S$, we assume that $u = \T(u)$ in the whole paper.
This generally holds for observation tables but \emph{not} for classification trees,
However, we can achieve it by extra checks and refinements later given in experiments section.

Now we show that, if the CEX returned to the TS learner is valid (cf.\ Definition~\ref{def:cex-gfr}), we can refine the current conjecture wDBA.
By analysing a valid CEX, we can add a new \emph{experiment} $e$ to the corresponding $E$ in order to distinguish two words $x\cdot a$ and $x'$ that are currently classified as equivalent, i.e., for $x, x'  \in S$ and $x \cdot a \in S \times \alphabet$, we have currently $f(x\cdot a) = f(x')$.

\begin{definition}\label{def:cex-gfr}
Let $\B = (\T, F)$ be the current conjecture wDBA and $(u, v)$ a decomposition.
We say $(u, v)$ is a \emph{valid} CEX of $\B$ if it satisfies that $\MQ(u, v) \neq \MQ(\Tilde{u}, v)$ where $\Tilde{u} = \T(u)$.
\end{definition}

In the following, we show that as long as $(u,v)$ is a valid CEX, we can use it to refine the current conjecture wDBA $\B$.

\paragraph{Refinement of $\B$.}
Since $\MQ(u, v) \neq \MQ(\Tilde{u}, v)$, it then holds that $u\not\canoEq \Tilde{u}$ because $v^{\omega}$ can be used to distinguish them.
We can find a new experiment and a new state as follows.
Let $k = |u|$ and $x_i = \T(\wordletter{u}{0\cdots i})$ be the state that $\B$ arrives in after reading the first $i$ letters of $u$.
Specially, $x_0 = \emptyword$ and $x_k = \Tilde{u}$.
We construct the following sequence via membership queries:
$\MQ(x_0 \cdot \wordletter{u}{0\ldots k}, v), \cdots, \MQ(x_i \cdot \wordletter{u}{i\ldots k}, v), \cdots, \MQ(\Tilde{u}\cdot \emptyword, v)$.
Since $\MQ(x_0\cdot u, v) \neq \MQ(\Tilde{u}, v)$ by assumption, this sequence must differ at some indices $0 \leq \ell, \ell + 1\leq k$.

That is, there must exist the smallest $\ell \in [0,k)$ such that $\MQ(x_{\ell} \cdot u[\ell] \cdot \wordletter{u}{\ell+1\ldots k}, v) \neq \MQ(x_{\ell+1} \cdot \wordletter{u}{\ell+1\ldots k}, v)$.
Then, since $x_{\ell+1} = \T(x_{\ell}\cdot u[\ell])$, we can use the experiment $e = (u[\ell + 1 \ldots k], v)$ to distinguish $x_{\ell} \cdot \wordletter{u}{\ell}$ and $x_{\ell+1}$.
Then, we add $x_{\ell}\cdot \wordletter{u}{\ell}$ to $S$ and $e$ to $E$.

Note that we use linear search to find $\ell$ above but we can also use binary search to further reduce the number of membership queries.
Therefore, upon receiving a valid CEX, the current conjectured wDBA will be refined.
\begin{lemma}\label{lem:make-progress}
The number of states in $\T$ will increase by one after a valid CEX $(u, v)$ is used to refine the current conjecture.   
\end{lemma}

\subsection{Acceptance marking}
\label{sec:acceptance-marking}
In this section, we will explain how to construct a weak DBA $\B = (\T, F)$ from the TS $\T$ by deciding the acceptance label for each state.
Since a run of a weak DBA eventually becomes trapped in either an accepting loop or a rejecting loop, it is quite natural to mark a state accepting if it belongs to the loop run of an accepting word and rejecting if belongs to the loop run of a rejecting word.

To this end, we call a state $u$ \emph{transient} if there exists no word $v \in \poswords$ such that $u = \T(uv)$. 
In other words, $u$ belongs to a trivial SCC. 
Conversely, if a state $u$ is in a non-trivial SCC, referred to as an \emph{SCC state}, there always exists a \emph{loop} word $v \in \poswords$ such that $u = \T(uv)$.

We will use a function $g: S \rightarrow \poswords$ to store a loop word for each state during the learning procedure.
This function $g$ is the core part of our algorithm as it allows us to easily analyse the counterexample received from the oracle as described in Section~\ref{sec:cex-analysis}.
More importantly, this function $g$ allows the counterexample analysis component to identify a valid counterexample for $\B$. 
Consequently, our learning algorithm can increment the number of states in the conjectured wDBA after each refinement guided by the counterexample.

\paragraph{Marking acceptance.}

Each \emph{transient} state $u$ is directly marked as \emph{rejecting}.  
For an \emph{SCC state} $u$, we first identify a loop word $v \in \poswords$ for $u$, set $g(u) = v$, and then mark $u$ as \emph{accepting} if $\MQ(u, v) = \top$, or as \emph{rejecting} otherwise.

Ideally, after the marking procedure, the resulting wDBA would have all states within the same SCC marked with the same acceptance label.  
In practice, however, as noted in~\cite{DBLP:journals/iandc/MalerP95}, conflicts may arise during the marking process if $\B$ is not yet the final automaton.  
One such conflict, %
referred to as C1, occurs when two states $u_1$ and $u_2$ within the same SCC have loop words $x$ and $y$, respectively, such that $u_1x^\omega \in L$ while $u_2y^\omega \notin L$.  

Since all states within the same SCC must be either entirely accepting or entirely rejecting, conflict C1 must be resolved before constructing the conjectured wDBA.  
To address this, we first introduce a second type of conflict, labelled C2, to which conflict C1 may be reduced: 
there is a state $u$ such that for two of its loop words $x, y \in \poswords$, it holds that $ux^{\omega} \in L$ but $uy^{\omega} \notin L$.
    
\begin{figure}
    \centering
     \scalebox{0.78}{
    \begin{tikzpicture}
    \node (p) at (11.2, 1) {};
    \node (q) at (11.2, 0) [draw, circle] {$u$};

    \node (pp0) at (9.6, 0.5) {$\top$};
    \node (pp1) at (12.8, 0.5) {$\bot$};
    \draw[->] (p) -- node[above] {} (q);
    \draw[loop left, ->, min distance=0.1cm, looseness=20] (q) to node[left] {$x$} (q);
    \draw[loop right, ->, min distance=0.1cm, looseness=20] (q) to node[right] {$y$} (q);

    \node (qq0) at (-1.8 + 4.6, 0.5) {$\top$};
    \node (qq1) at (1.8 + 6.6, 0.5) {$\bot$};
    \node (p1) at (4.6, 1) {};
    \node (q1) at (4.6, 0) [draw, circle] {$u_1$};
    \draw[->] (p1) -- node[above] {} (q1);
    \draw[loop left, ->, min distance=0.1cm, looseness=20] (q1) to node[left] {$x$} (q1);
    \node (p2) at (6.6, 1) {};
    \node (q2) at (6.6, 0) [draw, circle] {$u_2$};
    \draw[->] (p2) -- node[above] {} (q2);
    \draw[loop right, ->, min distance=0.1cm, looseness=20] (q2) to node[right] {$y$} (q2);

    \draw[->] (q1) to[bend left] node[above] {$z$} (q2);
    \draw[->] (q2) to[bend left] node[below] {$w$} (q1);
\end{tikzpicture}       %
     }
    \caption{Two conflict cases C1 (left) and C2 (right).}
    \label{fig:conflict-cases}
\end{figure}
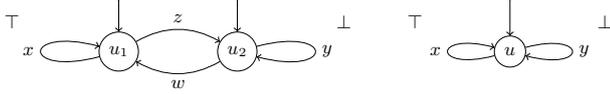

Figure~\ref{fig:conflict-cases} illustrates the two conflict cases.  
Our approach to resolving these conflicts is inspired by~\cite{DBLP:journals/iandc/MalerP95}.  
Specifically, we provide a concrete implementation of the conflict resolution framework proposed in~\cite{DBLP:journals/iandc/MalerP95} for cases C1 and C2.  
To resolve conflict C2, we invoke the function $\textsf{resolveConflict}(u, x, y)$, as described in Algorithm~\ref{alg:resolve-conflict}.

In order to make the explanation more general, we assume that $u = \T(ux) = \T(uy), ux^{\omega} \in L$ and $uy^{\omega} \notin L$.
We first explain the soundness of $\textsf{resolveConflict}(u, x, y)$, i.e., the returned counterexample is valid.
Since $u = \T(ux) = \T(uy)$, for any $h,k > 0$ and for $z=y^k\cdot x^k$, it follows that $u = \T(u \cdot z^{h-1}\cdot y^k) = \T(u \cdot z^h)$.  
In other words, both $u \cdot z^{h-1}\cdot y^k$ and $u \cdot z^h$ are identified as equivalent to $u$.

If $\MQ(u \cdot z^{h-1}\cdot y^k , x) = \bot$, it means that we can use $x^{\omega}$ to distinguish $u\cdot z^{h-1}\cdot y^k $ and $u$ since $ux^{\omega} \in L$.
Similarly, if $\MQ(u \cdot z^h , y) = \top$, we can then use $y^{\omega}$ to distinguish $u \cdot z^h $ and $u$ since $uy^{\omega} \notin L$.

For completeness, we show that the algorithm will terminate.
Assume that $\D$ is the minimal wDBA of the target language $L$ with $n$ states.
Since the number of SCCs in $\D$ is bounded by $n$, the word $m := u \cdot z^k$ for any $k \ge n$ may cause $\D$ to alternate between rejecting and accepting SCCs, but at most $n$ alternations.  
Beyond a certain point (i.e., as $k \ge n$), the words $u \cdot (y^n \cdot x^n)^k \cdot y^n$ and $u \cdot (y^n \cdot x^n)^k \cdot y^n \cdot x^n$ will eventually lead the run in $\D$ to be trapped in the same SCC, whether it is accepting or rejecting.

If the SCC is accepting, then $\MQ(m\cdot y^n \cdot x^n, y) = \top$, the algorithm will return on Line~10, otherwise the algorithm returns on Line~7.
Since $n$ is not known a priori, we increment $k$ in the outmost while loop, ensuring that $k$ will reach $n$ in the worst case.

\begin{algorithm}
\SetAlgoLined
\caption{$\textsf{resolveConflict}(u, x, y)$ where $u = \T_u(x) = \T_u(y)$, $u\cdot x^{\omega} \in L$ and $u\cdot y^{\omega} \notin L$ 
}\label{alg:resolve-conflict}
\KwIn{$u \in \finwords$ and $x, y \in \poswords$}
\KwOut{A valid counterexample}
\textsf{global }$k$ initialised to $1$ at the beginning\;
\While{true}{
    $h:=1$\;
    \While{$h \le k$}{
        $z:= y^k\cdot x^k$\;
        \If{$\MQ(u \cdot z^{h-1} \cdot y^k, x) = \bot$}{
            return $(u \cdot z^{h-1}  \cdot y^k, x) $ as a valid CEX\;
        }
        \If{$\MQ(u \cdot z^h, y) = \top$}{
            return $(u \cdot z^h, y) $ as a valid CEX\;
        }
        $h:=h+1$\;
    }
    $k : = k +1$\;
}
\end{algorithm}

We now discuss how to resolve conflict C1.
The idea is that we first find a word $z$ from state $u_1$ to $u_2$ and a word $w$ from $u_2$ to $u_1$.
This is possible since $u_1$ and $u_2$ are within the same SCC.
Clearly, $u_1 = \T_{u_1}(zw)$ and $u_2 = \T_{u_2}(wz)$.

We first check whether $\MQ(u_1, zw) = \bot$.
If so, there is a conflict C2 because $u_1 = \T_{u_1}(zw) = \T_{u_1}(x), ux^{\omega} \in L$ and $u_1\cdot (zw)^{\omega} \notin L$.
We can then call $\textsf{resolveConflict}(u_1, x, zw)$ to obtain a valid CEX.
Similarly, if $\MQ(u_2, wz) = \top$, we find a conflict C2 because $u_2 = \T_{u_2}(wz) = \T_{u_2}(y), u_2(wz)^{\omega} \in L$ and $u_2\cdot (y)^{\omega} \notin L$.
We can also call $\textsf{resolveConflict}(u_2, wz, y)$ to obtain a valid CEX.
In the case of $\MQ(u_2, wz) = \bot$ and $\MQ(u_1, zw) = \top$ (or equivalently $\MQ(u_1z, wz) = \top$), it then follows that $(wz)^{\omega}$ can distinguish $u_2$ and $u_1z$ since $u_2 = \T(u_1z)$.
Hence, $(u_1z, wz)$ is a valid CEX.

\begin{algorithm}[htb]
\SetAlgoLined
\caption{$\textsf{resolveConflictState}(u_1, u_2, x, y)$ where $u_1 = \T_{u_1}(x), u_2 = \T_{u_2}(y), u_1\cdot x^{\omega} \in L$ and $u_2\cdot y^{\omega} \notin L$
}\label{alg:resolve-conflict-state}
\KwIn{$u_1, u_2 \in \finwords$, $x, y \in \poswords$}
\KwOut{A valid counterexample}
Let $z \in \poswords$ such that $u_2 = \T_{u_1}(z)$ \;
Let $w \in \poswords$ such that $u_1 = \T_{u_2}(w)$ \;
\If{$\MQ(u_1, zw) = \bot$}{
            return $\textsf{resolveConflict}(u_1, x, zw) $ \;
}
\If{$\MQ(u_2, wz) = \top$}{
            return $\textsf{resolveConflict}(u_2, wz, y) $ \;
}        
return $(u_1z, wz) $ as a valid CEX\;
\end{algorithm}

This then completes the construction of the conjectured wDBA $\B$ and we can ask an equivalence query $\EQ(\B)$.

We remark that the acceptance marking procedure in~\cite{DBLP:journals/iandc/MalerP95} works by traversing all rows $u$ and columns $(x, y)$ in the observation table and mark the states in the loop run over $uxy^{\omega}$ according to its entry value.
It may happen that some SCCs are not marked because they are not covered by the words in the observation table.
These SCCs are marked arbitrarily in their algorithm and must be fixed by counterexamples returned from the oracle, while our marking algorithm guarantees that every SCC state will be marked according to their real loop words.

\subsection{Counterexample analysis}
\label{sec:cex-analysis}

When the conjectured wDBA $\B$ is not correct, the teacher will reply with the answer ``no" alongside a CEX $w \in \upword{L} \ominus \lang{\B}$, represented by a decomposition. %
In this section, we will make use of the CEX $w$ to refine the current conjecture $\B$.
We say the CEX $w$ is \emph{negative} if $w \in \lang{\B}\setminus L$ and \emph{positive} if $w \in L \setminus \lang{\B}$.
As usual, since $\B = (\T, F)$ has a finite number of states, we can easily obtain a normalised decomposition $(x, y)$ of $w$ such that $\T(x) = \T(xy)$.
Let $u = \T(x) = \T_{u}(y)$.

We first analyse the case when $w$ is a negative counterexample, i.e., $w \in  \lang{\B}\setminus L$.
This means that the run $\run$ over $w$ gets trapped in an accepting SCC $C$ of $\B$, but it should not be accepted.
Since $u$ is in an accepting SCC and $\T$ is a conflict-free TS, we can find its loop word $v = g(u)$ such that $u = \T_u(v)$ and $uv^{\omega} \in L$.
Recall that in the construction of wDBA, we use function $g$ to record the loop word for each SCC state.
We first test whether $\MQ(u, y) = \top$, i.e., whether $uy^{\omega}$ is in $L$ or not.
If $uy^{\omega} \in L$, we can use $y^{\omega}$ to distinguish $x$ and $u$ since $xy^{\omega} \notin L$ and $x$ is currently classified as being equivalent to $u$.
We then return $(x, y)$ as a valid CEX to refine $\T$.
Otherwise $uy^{\omega} \notin L$, together with the fact that $u = \T_u(v), uv^{\omega} \in L, u = \T_{u}(y)$ and $uy^{\omega} \notin L$, we find a conflict type C2.
We can then call $\textsf{resolveConflict}(u, v, y)$ to obtain a valid counterexample.

Now we can analyse the case when $w$ is a positive counterexample, i.e., $w \in L \setminus \lang{\B}$.
Again, this indicates that the run $\rho$ over $w$ is trapped in a rejecting SCC $C$ but it should be accepted.
Since $C$ is rejecting, we can find the loop word $v = g(u)$ such that $u = \T_u(v)$ and $uv^{\omega} \notin L$.
We can test whether $\MQ(u, y) = \top$, i.e., whether $uy^{\omega}$ is in $L$ or not.
Again, if $uy^{\omega} \notin L$, we can then use $y^{\omega} $ to distinguish $u$ and $x$ because $u = \T(x)$, indicating that $x$ is currently equivalent to $u$.
Then, we can return $(x, y)$ as a valid CEX to refine $\B$.
Otherwise $uy^{\omega} \in L$, this then leads to a conflict type C2 because $u = \T_{u}(y), u = \T_u(v), uy^{\omega} \in L$ and $uv^{\omega} \notin L$.
Hence, we can call $\textsf{resolveConflict}(u, y, v)$ to obtain a valid CEX to refine $\B$.
\begin{figure}
    \centering
    \includegraphics[width=0.9\linewidth]{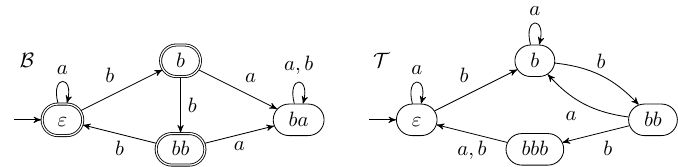}
    \caption{ 
     (a) Left: A wDBA $\B$ constructed from the conflict-free TS $\T$, with the mapping $g = \{\emptyword \mapsto a, b \mapsto bbb, ba \mapsto a, bb \mapsto bbb\}$.  
     (b) Right: A TS $\T$ with the mapping $g = \{\emptyword \mapsto a, b \mapsto a, bb \mapsto ab, bbb \mapsto abbb\}$, exhibiting conflicts.
     }
    \label{fig:ts-wdba-example}
\end{figure}
Lemma~\ref{lem:cex-make-progress} then follows.
\begin{lemma}\label{lem:cex-make-progress}
A valid CEX will be constructed to refine the current wDBA once a counterexample is returned by the oracle.   
\end{lemma}

\begin{example}
    (a) {\bf Conflict-free TS} Figure~\ref{fig:ts-wdba-example}(a) depicts a TS $\T$ with the function $g$ and its corresponding wDBA $\B = (\T, F)$ where $F = \{\emptyword, b, bb\}$ during the learning procedure of $\D$ in Figure~\ref{fig:wdba-example}.
    We can see that for every accepting state $u \in F$, we have $u\cdot (g(u))^{\omega} \in \lang{\D}$, while for the rejecting state $ba$, it holds that $ba\cdot (g(ba))^{\omega} = ba\cdot a^{\omega} \notin L$.
    In this case, $\T$ is a conflict-free TS with respect to the function $g$.

    Thus, we pose the equivalence query $\EQ(\B)$. 
    Since $\lang{\D} \neq \lang{\B}$, the oracle responds with the answer ``no" along with a counterexample $w \in \lang{\D} \ominus \lang{\B}$. 
    Suppose $w = bbbba a^\omega$, which is a positive counterexample such that $w \in \lang{\D}$ and $w \notin \lang{\B}$.
    We then obtain a normalized decomposition $(x, y) = (bbbba, a)$ of $w$. 
    Here, $\T(x) = \T(xy)$, and we have $\T(bbbba) = ba$. 
    Since $ba \cdot y^\omega \notin \lang{\D}$ but $x \cdot y^\omega \in \lang{\D}$, we can use $y^\omega = a^\omega$ to distinguish $ba$ and $x = bbbba$. 
    Consequently, $(bbbba, a)$ is returned as a valid CEX to the TS learner to refine $\B$.

    To refine $\B$, we examine the state $ba = \T(bbbba)$ and the corresponding sequence of membership queries:  
    \begin{align*}
        &\MQ(\epsilon \cdot bbbba, a), \MQ(b \cdot bbba, a), \MQ(bb \cdot bba, a), \\ &\MQ(\epsilon \cdot ba, a), \MQ(b \cdot a, a), \MQ(ba \cdot \epsilon, a).
    \end{align*}

    The first position at which the query results differ is $\MQ(bb \cdot bba, a) = \top$ and $\MQ(\epsilon \cdot ba, a) = \bot$. 
    Based on this discrepancy, we identify the experiment $e = (ba, a)$, which serves to distinguish between $bb \cdot b$ and $\epsilon$. 
    The word $bb \cdot b$ is then added to $S$.
    This refinement ensures the correct TS for $\D$, and with the correct acceptance labelling of $ba$, the wDBA is accurately constructed.

    (b) {\bf TS with conflicts}
    If the TS $\T$ is not conflict-free—for instance, as in the TS depicted in Figure~\ref{fig:ts-wdba-example}(b)—we first resolve the conflict before posing an equivalence query. In this case, $\epsilon \cdot (g(\epsilon))^{\omega} \in L$, whereas $b \cdot (g(b))^{\omega} = b \cdot a^{\omega} \notin L$. To address this, we invoke $\textsf{resolveConflictState}(\epsilon, b, a, a)$.
    
    We set $z = b$ and $w = bba$, where $\epsilon = \T_b(w)$ and $b = \T_\epsilon(z)$. Following Algorithm~\ref{alg:resolve-conflict-state}, we first check $\MQ(\epsilon, zw) = \MQ(\epsilon, bbba) = \bot$. The response is positive since $\epsilon \cdot (bbba)^{\omega} \in \lang{\D}$. We then test $\MQ(b, wz) = \MQ(b, bba \cdot b) = \top$, which is also confirmed.
    
    At this point, we call $\textsf{resolveConflict}(b, bbab, a)$ as per Algorithm~\ref{alg:resolve-conflict}.
    The algorithm terminates at $k = 1$ and $h = 1$ at line~7 because $\MQ(m \cdot y, x) = \bot$, where $m = u = b$, $y = a$, and $x = bbab$. This produces a valid counterexample $(ba, bbab)$ for the TS learner.
    
    The TS learner uses this counterexample to identify an experiment $(\epsilon, bbab)$, distinguishing $b$ from $ba$. Consequently, $ba$ is added to $S$, refining the TS of $\D$. With the correct acceptance labelling of $ba$, this process yields the correct wDBA.
    \end{example}

\subsection{Correctness and Complexity}
Let $\D$ be the target minimal wDBA of $L$ and $n$ the number of states in $\D$.
Three aspects of correctness and complexity are easy to establish:

\begin{enumerate}
    \item the states in the successively built function $f$ have pairwise different right languages -- this entails that we can produce at most $n$ states;
    \item one state is added after every failed equivalence query -- thus, the number of equivalence queries is bound by $n$;
    \item we only add one experiment per failed equivalence query -- the size of function $f$, and the number of membership queries used to build it, is therefore in $\bigO(n^2 \cdot|\Sigma|)$.
\end{enumerate}

The remaining part of complexity is the cost of identifying the experiments added to the table.
They fall into failed and successful inner loops of Algorithm \ref{alg:resolve-conflict}, both of which occur at most $n$ times and require $\bigO(n^2)$ many membership queries; a further $\bigO(n^2)$ membership queries to determine acceptance marking for the SCCs, and finally the membership queries to identify the exact experiments for refining $\T$.

The latter depend on the length $\ell$ of a counterexample returned by a failed equivalence query; more precisely, the length of the words that we have to base the search on can be up to $\bigO(n^3 + n^2 \ell)$.
However, we note that this word can be traversed with a logarithmic search as described in~\cite{DBLP:journals/iandc/RivestS93}, so that only $\bigO(\log(n^3 + n^2 \ell))$ membership queries of this type are needed for each of the up to $n$ failed equivalence queries.
Thus, we require $\bigO(\max(n^2\cdot |\alphabet|, n \cdot \log \ell))$ membership queries, as $\bigO(n\cdot \log (n^2\cdot \ell + n^3))=\bigO(n\cdot \log (n^2\cdot (\ell + n)))=\bigO(n \cdot \log \ell)$ holds when $\ell \in 2^{\bigO(n\cdot |\alphabet|)}$; 
see the appendix for more details.

\begin{restatable}{theorem}{thmMain}

Our wDBA learning algorithm needs at most $n$ equivalence queries and $\bigO(\max\{n^2\cdot |\alphabet| , n\cdot \log \ell\})$ membership queries.

Our wDBA learning algorithm is correct and runs in time polynomial in $n$. 
\end{restatable}

In~\cite{DBLP:journals/iandc/MalerP95}, the guarantee provided is $\bigO(n^3 \cdot \ell \cdot |\Sigma|)$; with the additional assumption that the counterexamples provided are minimal, and thus $\ell \in \bigO(n^2)$, this results in $\bigO(n^5 \cdot |\Sigma|)$; note that we get our $\bigO(n^2 \cdot |\Sigma|)$ guarantee under the assumption that $\ell \in 2^{\bigO(n \cdot |\Sigma|)}$, a much milder assumption than minimality.

We note that as all our conjecture automata are weak by construction,
the algorithm would not terminate if the target language is
not weak---just as $L^*$~\cite{DBLP:journals/iandc/Angluin87} would not terminate if learning, say, a
pushdown language.

\begin{figure}[htb]
    \centering
    \includegraphics[width=1\linewidth]{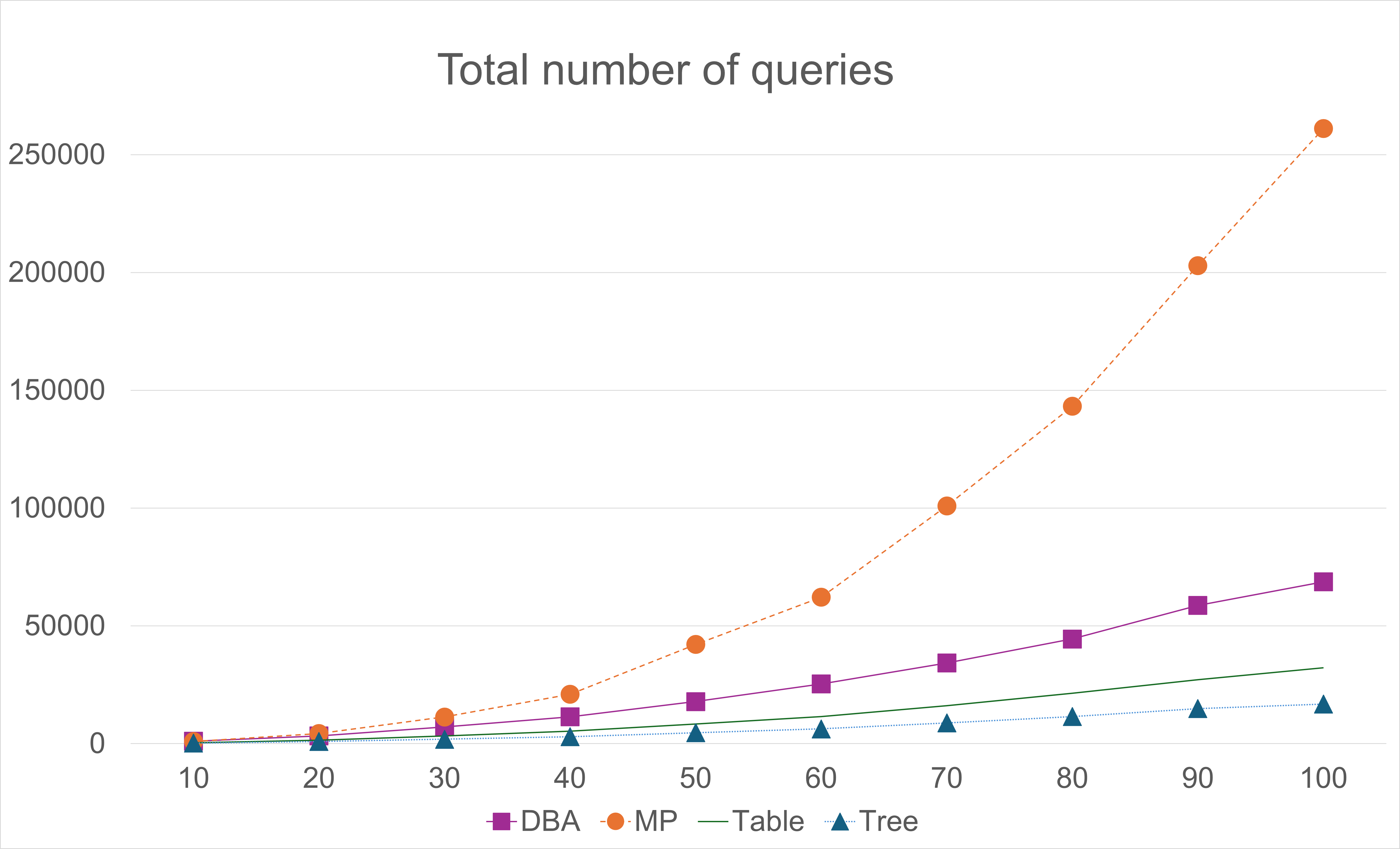}
    \caption{Total number of queries, the sum of equivalence queries and membership queries, for learning wDBAs.}
    \label{fig:queries}
\end{figure}
\begin{figure}[htb]
    \centering
    \includegraphics[width=1\linewidth]{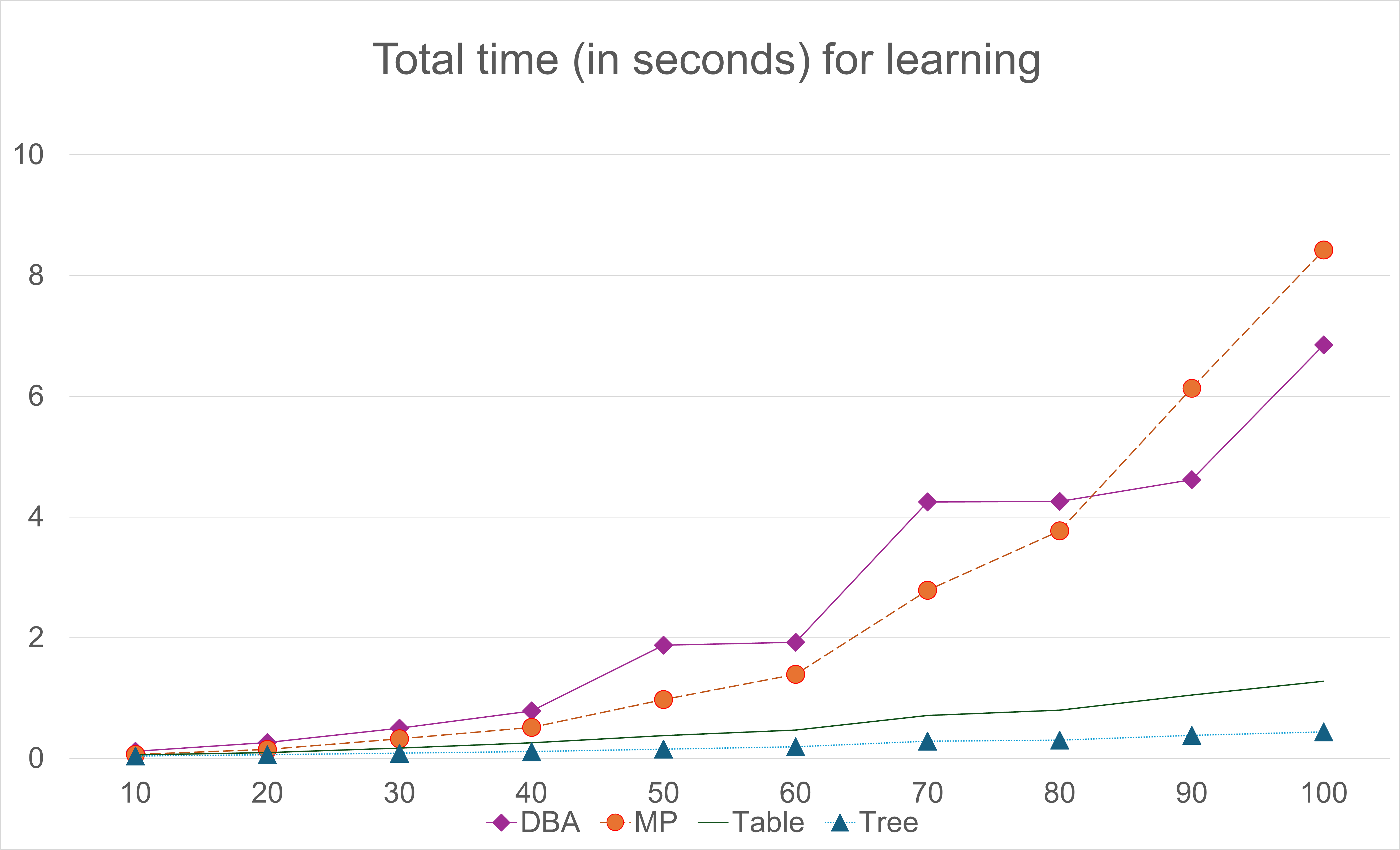}
    \caption{Total time (in seconds) for learning wDBAs.}
    \label{fig:tto}
\end{figure}

\section{Experiments}
\label{sec:experiments}
\ly{Mona: the paper needs to be anonymous, which means that every thing needs to be written in third person. In particular, we cannot say something like our DBA learning algorithm, or our ROLL library. Because now we should act like a third person when writing the paper.}
To demonstrate the efficiency of our wDBA learning algorithm, we compare it with the algorithm proposed by %
\cite{DBLP:journals/iandc/MalerP95}, referred to as \textsf{MP} and the DBA learning algorithm from~\cite{LiST24}, referred to as \textsf{DBA}.
Recall that a DBA is learned in~\cite{LiST24} through learning limit FDFAs.
Our algorithm is implemented using two data structures: observation tables (\textsf{Table}) and classification trees (\textsf{Tree}).
Note that, for the tree-based approach, we may have that for a state $u \in S$, $u \neq \T(u)$.
Let $u' = \T(u)$. Since $u$ and $u'$ are distinguished by some experiment $e=vy^{\omega} \in E$, we can return the example $(uv,y)$ to refine the TS $\T$.
This is because $\MQ(uv, y) \neq \MQ(u'v, y)$, so there will be different membership query results between the first $\MQ(x_0 \cdot uv, y)$ and a middle $\MQ(x_{|u|}\cdot v, y)$ where $x_0 = \emptyword$ and $x_{|u|} = u'$ are two state representatives in the sequence when refining $\T$.
\revise{We note that there are more advanced techniques to make sure that $u = \T(u)$ holds, see e.g.~\cite{DBLP:conf/rv/IsbernerHS14}. }
\iffalse
We also included the DBA learning algorithm from~\cite{LiST24} in our comparison, which learns a DBA by learning a family of DFAs.

However, since their approach targets general DBAs rather than wDBAs, it performed significantly worse on our benchmarks. 
\revise{As a result, we only include it in the appendix.}
\fi

We have randomly generated 500 minimal wDBAs with state sizes ranging from 10 to 100, increasing in steps of 10. 
For each state size, we have generated 50 wDBAs.
\revise{
These wDBAs are non-trivial, each containing between 2 and 10 non-trivial SCCs.
A non-trivial SCC is one with at least two states.
We would like to point out that passing the equivalence query guarantees that the learned wDBA and the target wDBA are language equivalent.
This implies that all our learned automata are correct.}

To evaluate the performance of each algorithm, we measured the average number of queries required to learn the target wDBA and the average running time. As shown in Figure~\ref{fig:queries}, \textsf{Tree} consistently requires the fewest queries among all algorithms, followed by \textsf{Table}, \textsf{DBA}, and, lastly, \textsf{MP}. 
The advantage of our algorithms over \textsf{MP} and \textsf{DBA} becomes more pronounced as the size of the wDBA increases. Both the average number of queries and the time taken to learn a wDBA (cf. Figure~\ref{fig:tto}) grow significantly more slowly with automaton size for \textsf{Table} and \textsf{Tree} compared to \textsf{MP} and \textsf{DBA}. All algorithms are available for use in the learning library ROLL\footnote{Available at \url{https://github.com/iscas-tis/roll-library}}~\cite{rolllibrary}. 
We provide an artifact \cite{artifact_wDBA_Learning_2025} to reproduce all experimental results reported in this paper.
 
\revise{
We are aware of a collected benchmark in~\cite{DBLP:conf/birthday/NeiderSVK97}\footnote{Available at \url{https://automata.cs.ru.nl/}}, which contains DFAs, Moore machines, Mealy machines, interface automata and register automata. 
For our experiments, we focus on the DFAs. 
The benchmark provides two types of DFAs: one with 1,000 states and the other with 2,000 states. 
Each DFA contains a single non-trivial SCC. 
To construct wDBAs, we applied a simple transformation by making all states within the SCC as accepting. 
The resulting wDBAs preserve the original structure, yet their minimal equivalents reduce to a single-state automaton. 
Our algorithms, \textsf{Table} and \textsf{Tree}, consistently require significantly fewer queries than \textsf{MP} to learn these wDBAs.
Additional details regarding the results can be found in the appendix.
We do not see an easy way to use the other benchmarks in our experiments and thus do not include them in the evaluation.
}

\section{Conclusion}
\label{sec:conclusion}
We have introduced a novel learner for weak languages.
It improves over the state-of-the-art in terms of complexity, where we reduce the number of queries from quintic to quadratic, with milder assumptions on the length of the counterexamples provided by equivalence queries -- where the old quintic result relies on them being quadratic in the size of the automaton, while we can allow for it to be exponential.
Further, we have formulated our framework such that it works more flexibly with other data structures, like the classification trees we have implemented.
Our experimental results show that, irrespective of the data structure chosen, we always outperform the state-of-the-art algorithm from \cite{DBLP:journals/iandc/MalerP95}.

From a practical point of view, the advantage our algorithms holds over \cite{DBLP:journals/iandc/MalerP95} seems to grow with the size of the automaton.
Even for the small 100 state automata, the largest automata we used for evaluation, the advantage our table-based implementation holds over \cite{DBLP:journals/iandc/MalerP95} is almost an order of magnitude, while moving to the more modern classification trees as the underpinning data structure further halves the number of queries.
With better efficiency, our work would further benefit the applications of learning algorithms in various fields, including verification, testing, and understanding of ML models.

\newpage

\paragraph{Acknowledgements.}
This work was supported in part by the CAS Project for Young Scientists in Basic Research (Grant No. YSBR-040), ISCAS Basic Research (Grant Nos. ISCAS-JCZD-202406, ISCAS-JCZD-202302), ISCAS New Cultivation Project ISCAS-PYFX-202201, and by the EPSRC through grants EP/X03688X/1 and EP/X042596/1.

\appendix

\clearpage{}%
\section{Correctness and Complexity}\label{appendix}
Let $\D$ be the target minimal wDBA of $L$ and $n$ the number of states in $\D$.
Three aspects of correctness and complexity are easy to establish:

\begin{enumerate}
    \item the states in the successively built function $f$ have pairwise different right languages -- this entails that we can produce at most $n$ states;
    \item one state is added after every failed equivalence query -- thus, the number of equivalence queries is bound by $n$;
    \item we only add one experiment per failed equivalence query -- the size of the function $f$, and the number of membership queries used to build it, is therefore in $\bigO(n^2 |\Sigma|)$.
\end{enumerate}
\vspace{-.5em}
The remaining part of complexity is the cost of identifying the experiments added to the table.

Since in Algorithm~\ref{alg:resolve-conflict}, the variable $k$ is a global variable and only increases to at most to $n$ during learning; these increases happen when the inner loop of Algorithm \ref{alg:resolve-conflict} was unsuccessful in identifying a counterexample.
The number of membership queries resulting from these unsuccessful attempts is $\bigO(n^2)$ and thus dominated by the $\bigO(n^2 \cdot |\Sigma|)$ queries from (3).

When the inner loop is successful in identifying a counterexample, a new state is added, and this happens only $\leq n$ times, so that the number of membership queries used in successful iterations of the inner loop is bounded by $2n^2$.

For each acceptance marking, we only use $t$ membership queries for a TS with $t$ states to decide the acceptance of each state.
The number of membership queries is $\sum_{t=1}^n t \in \bigO(n^2)$.
In the whole learning procedure, Algorithm~\ref{alg:resolve-conflict-state} uses $2n$ membership queries because it can only be called at most $n$ times.

Now we discuss about the number of membership queries used in refining $\T$.
For a valid CEX $(u', v')$ for $\T$, the number of membership queries used is $\bigO(|u'|)$ if we use linear search and $\bigO(\log |u'|)$ when we use binary search.
So we only need to estimate the length of $u'$.

First, when refining the TS $\T$, a new state representative $u\cdot a$ is obtained from an existing state $u$ and a letter extension $a$.
Hence, the state representatives in the correct $\T$ has length at most $n$ since the first state representative $\emptyword$ has length $0$ and there are only $n$ state representatives.

There are two sources of counterexamples for refining $\T$:
one from acceptance marking and the other from CEX analysis component.
For the first, the input words $u, x$ and $y$ have length $\bigO(n)$, where $x$ and $y$ are two different loop words of state $u$.
So, the length of the prefix $u'$ returned from two algorithms will have length $\bigO(n^3)$.

Now we discuss the length of prefixes of counterexamples generated by the counterexample analysis component when a CEX $(x', y')$ is returned from the oracle.
Let $\ell = |x'| + |y'|$ be the length of the \emph{longest} CEX returned.

The length of the prefix $u'$ of a valid counterexample $u\cdot [y^k\cdot x^k]^h$ or $u\cdot [y^k\cdot x^k]^{h-1}\cdot y^k$ returned by \textsf{resolveConflict} is $\bigO(n^2\cdot \ell + n^3)$, where $|u|$ is bounded by $n$ and, for $|x|$ and $|y|$, one is bounded by $\ell$ and the other by $n$ in the worst case.

Therefore, we only need  $\bigO(n^2\cdot \ell + n^3)$ membership queries for refining conjecture TS $\T$ every time if we use linear search.
But we only need $\bigO(\log (n^2\cdot \ell + n^3))$ membership queries with binary search.
Since we need to refine $\T$ for at most $n$ times, then the number of membership queries needed there is $\bigO(n\cdot \log (n^2\cdot \ell + n^3))$.
It then follows that the number of membership queries consumed by our algorithm is $\bigO(\max(n^2\cdot |\alphabet|, n\cdot \log (n^2\cdot \ell + n^3))$.

We note that $\bigO(n\cdot \log (n^2\cdot \ell + n^3)) = \bigO(n\cdot \log (n^2\cdot (\ell + n))) = \bigO(n\cdot (\log n^2 + \log (\ell + n)))=\bigO(n \cdot \log \ell)$ holds if the second term dominates and we assume that $\ell \in 2^{\bigO(n\cdot |\alphabet|)}$.
So in summary, we obtain our main result.
\ly{If $\ell$ is polynomial in $n$, then we have $\bigO(n^2\cdot |\alphabet|)$.}
\sven{Yes, but actually: if $\ell \in 2^{\bigO(n\cdot |\Sigma|)}$, we have $\bigO(n\cdot |\Sigma|)$.}
\ly{Excellent. I think this is very reasonable assumption. If the teacher is nice, she should be able to provide a counterexample in $\bigO(n^2)$. But of course $\ell \in 2^{\bigO(n\cdot |\Sigma|)}$ is also fine.}

\thmMain*

In~\cite{DBLP:journals/iandc/MalerP95}, the guarantee provided is $\bigO(n^3 \cdot \ell \cdot |\Sigma|)$; with the additional assumption that the counterexamples provided are minimal, and thus $\ell \in \bigO(n^2)$, this results in $\bigO(n^5 \cdot |\Sigma|)$; note that we get our $\bigO(n^2 \cdot |\Sigma|)$ guarantee under the assumption that $\ell \in 2^{\bigO(n \cdot |\Sigma|)}$, a much milder assumption than minimality.

\section{Experiments}\label{app:exp}
{\bf DBA Learning Algorithm}
We present the complete experimental results on the randomly generated wDBAs with the learning algorithm for DBAs proposed in~\cite{LiST24}, which is referred to as {\sf DBA}.
The table below reports the average number of queries required to learn the target wDBAs by the four algorithms.
\begin{table}[h!]
\centering
\resizebox{.72\linewidth}{!}{%
\begin{tabular}{|c|c|c|c|c|}
\hline
\rowcolor{gray!10}
\textbf{Size} & {\sf DBA} & {\sf MP} & {\sf Table} & {\sf Tree} \\
\hline
10   & 1022.70   & 960.64     & 393.22     & 254.46    \\
20   & 3327.54   & 4283.12    & 1454.60    & 900.94    \\
30   & 7100.54   & 11337.86   & 3262.88    & 1886.54   \\
40   & 11418.94  & 20932.16   & 5265.00    & 2936.94   \\
50   & 17869.08  & 42094.92   & 8317.58    & 4630.58   \\
60   & 25349.06  & 62179.50   & 11510.66   & 6276.00   \\
70   & 34230.96  & 100934.16  & 16083.08   & 8810.94   \\
80   & 44437.68  & 143230.18  & 21410.68   & 11513.10  \\
90   & 58640.84  & 202896.64  & 27128.64   & 14841.84  \\
100  & 68700.80  & 261111.70  & 32286.96   & 16813.64 \\
\hline
\end{tabular}
}
\label{tab:fdfa1}
\end{table}

The table below presents the average running time required by the four algorithms to learn the target wDBAs.

\iffalse
\begin{table}[h!]
\centering
\resizebox{.72\linewidth}{!}{%
\begin{tabular}{|c|c|c|c|c|}
\hline
\rowcolor{gray!10}
\textbf{Size} & {\sf DBA} & {\sf MP} & {\sf Table} & {\sf Tree} \\
\hline
10  & 0.64046  & 0.20248  & 0.18550  & 0.13108 \\
20  & 0.68154  & 0.33880  & 0.27734  & 0.20706 \\
30  & 0.99424  & 0.62396  & 0.38474  & 0.25604 \\
40  & 1.47704  & 0.96962  & 0.52522  & 0.29782 \\
50  & 2.72996  & 1.71700  & 0.70730  & 0.37126 \\
60  & 2.93506  & 2.31848  & 0.88414  & 0.44024 \\
70  & 5.29080  & 3.82908  & 1.20850  & 0.55768 \\
80  & 5.94080  & 4.99282  & 1.52908  & 0.64500 \\
90  & 6.68306  & 7.44136  & 1.99838  & 0.79854 \\
100 & 10.53984 & 10.43978 & 2.30561  & 0.88788 \\
\hline
\end{tabular}
}
\label{tab:fdfa2}
\end{table}

\fi

\begin{table}[h!]
\centering
\resizebox{.72\linewidth}{!}{%
\begin{tabular}{|c|c|c|c|c|}
\hline
\rowcolor{gray!10}
\textbf{Size} & {\sf DBA} & {\sf MP} & {\sf Table} & {\sf Tree} \\
\hline
10  & 0.11780  & 0.06510  & 0.05724  & 0.04030 \\
20  & 0.26428  & 0.14988  & 0.09938  & 0.06324 \\
30  & 0.50000  & 0.32686  & 0.17346  & 0.08648 \\
40  & 0.78704  & 0.51622  & 0.26170  & 0.11286 \\
50  & 1.87838  & 0.97790  & 0.37732  & 0.15364 \\
60  & 1.92722  & 1.39432  & 0.46960  & 0.19368 \\
70  & 4.24870  & 2.78708  & 0.71104  & 0.28710 \\
80  & 4.25622  & 3.76858  & 0.80084  & 0.30426 \\
90  & 4.61958  & 6.13236  & 1.04892  & 0.38436 \\
100 & 6.85212  & 8.42328  & 1.27996  & 0.44038 \\
\hline
\end{tabular}
}
\label{tab:fdfa2}
\end{table}

{\bf DFA Benchmarks} 
We present experimental results on the DFA benchmarks introduced in~\cite{DBLP:conf/birthday/NeiderSVK97}. 
After transformation and minimisation, all DFAs with 1,000 states result in trivial wDBAs—each comprising a single state with the same number of transitions. 
The same outcome is observed for DFAs with 2,000 states. 
Thus, although there are 3,600 DFAs of each type, the number of queries required by each algorithm remains consistent across all instances of the same type. 
For both sets of wDBAs (prior to minimisation), our algorithms consistently outperform {\sf MP}, while {\sf DBA} fails to handle these cases due to stack overflows, as shown in the table below.%

\begin{table}[htb]
\centering
\resizebox{.74\linewidth}{!}{%
\renewcommand{\arraystretch}{1.5}
\begin{tabular}{l|rrrc|rrr}
\toprule
\rowcolor{gray!10}
\textbf{Method} & \textbf{EQs} & \textbf{MQs} & \textbf{Total} & \hspace{1em} & \textbf{EQs} & \textbf{MQs} & \textbf{Total} \\
\rowcolor{gray!10}
& \multicolumn{3}{c}{\textbf{1,000 states}} & & \multicolumn{3}{c}{\textbf{2,000 states}} \\
\midrule
{\sf DBA}    & - & - & - & & - & -  & - \\
{\sf MP}    & 1 & 420 & 421 & & 1 & 110 & 111 \\
{\sf Table} & 1 &  22 &  23 & & 1 &  12 &  13 \\
{\sf Tree}  & 1 &   1 &   2 & & 1 &   1 &   2 \\
\bottomrule
\end{tabular}
}
\label{tab:dfa}
\end{table}

\clearpage{}%

\end{document}